# Multi-Bit Read and Write Methodologies for Diode-STTRAM Crossbar Array

Mohammad Nasim Imtiaz Khan, Swaroop Ghosh, Radha Krishna Aluru and *Rashmi Jha
Computer Science and Engineering, University of South Florida
*University of Cincinnati
{khan12, swaroopghosh, aluru}@mail.usf.edu, jhari@ucmail.uc.edu

*Abstract— Crossbar arrays using emerging non-volatile memory technologies such as Resistive RAM (ReRAM) offer high density, fast access speed and low-power. However the bandwidth of the crossbar is limited to single-bit read/write per access to avoid selection of undesirable bits. We propose a technique to perform multi-bit read and write in a diode-STTRAM (Spin Transfer Torque RAM) crossbar array. Simulation shows that the biasing voltage of half-selected cells can be adjusted to improve the sense margin during read and thus reduce the sneak path through the half-selected cells. In write operation, the half-selected cells are biased with a pulse voltage source which increases the write latency of these cells and enables to write 2-bits while keeping the half-selected bits undisturbed. Simulation results indicate biasing the half-selected cells by 700mV can enable reading as much as 512-bits while sustaining 512x512 crossbar with 2.04 years retention. The 2-bit writing requires pulsing by 50mV to optimize energy.*

## I. INTRODUCTION

Spin Transfer Torque RAM (STTRAM) with Magnetic Tunnel Junction (MTJ) as storage element offers multitude of advantages including non-volatility, speed, scalability, endurance and low-power. However, the footprint is limited by three-terminal transistor-based selectors. Selector diodes (SD) such as Mixed Ionic Electronic Conduction (MIEC) [1], Metal-Insulator-Metal (MIM) tunneling [2], oxide heterojunction p-n, and Schottky diodes [3] for resistive RAM (ReRAM) have been extensively studied [4]. Integration of SD and ReRAM have been attempted due to material compatibility and simplistic structure. Unlike ReRAM, MTJ do not need electroforming which simplifies the integration however, new challenges are imposed such as low TMR (Tunnel Magneto-Resistance), and, read/write and retention sensitivity to the noise and variations. In [5], a Metal-Insulator- Insulator-Metal (MIIM) SD is proposed to enable 3D integration with STTRAM (Fig. 1). Although SD increases the switching voltage which is undesirable for cache it allows high-density and scalability by enabling 3D stacking for new applications e.g., Internet-of-Things.

Several emerging non-volatile memory technologies have been proposed for crossbar array [6-9]. Techniques such as V/2 and V/3 read/write have been studied to reduce sneak path current in crossbar arrays [10]. However, the crossbar array is bit addressable as shown in Fig. 1 (a) to avoid disturbing the half-selected and unselected cells which limits the bandwidth. Furthermore it also increases the number of crossbars that need to be activated to access a cache line increasing power dissipation from peripherals and extra sneak path leakage. Multi-bit write for ReRAM crossbar has been proposed in [7] [11] where the '1's are written in $1^{st}$ cycle and '0's are written in the $2^{nd}$ cycle. This in turn increases the write latency. Multi-level cell (MLC) STTRAM has been proposed that can store 2 bit per cell and thus ensure multi-bit read-write [12] [13]. However, the write operation becomes complicated and the read operation requires 4 senseamps per cell to distinguish 4 states of 2 bits increasing the circuit complexity.

In this paper we propose techniques to perform multi-bit read and write operation on the STTRAM crossbar (Fig. 1(b)). The selector device proposed in [5] is employed and optimized to make trade-off among energy, speed, retention time and sense margin. By performing 2-bit read and write operation per cycle we reduce the number of active crossbars per access by half. The read operation exploits the effect of biasing voltage on the half selected cells to increase the sense margin. In read operation instead of V/2 scheme, V/2 + Δ (where Δ is a positive value) is proposed which reduces the sneak path current and improves the sense margin. The write operation exploits the nonlinear dependence of write latency of STTRAM on write voltage to write 2-bits while keeping the half-selected bits undisturbed. Instead of biasing the half-selected cells with a dc voltage, a pulse voltage source is used which increases the effective write latency of the half-selected cells and gives enough time to write two selected bits simultaneously.

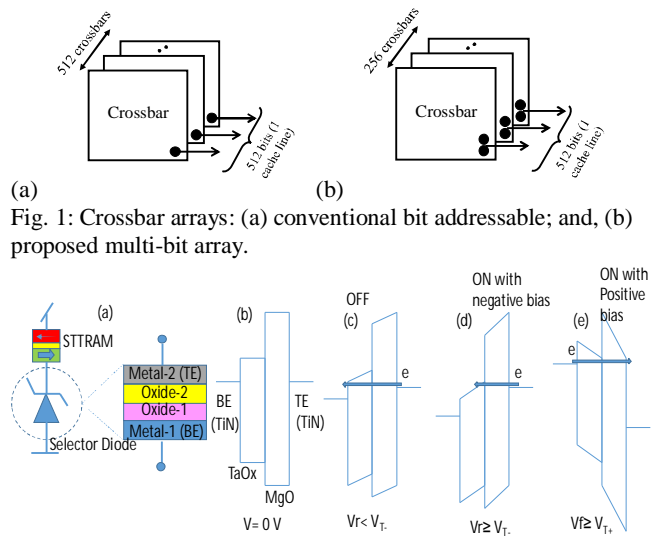

(a)                 (b)

Fig. 1: Crossbar arrays: (a) conventional bit addressable; and, (b) proposed multi-bit array.

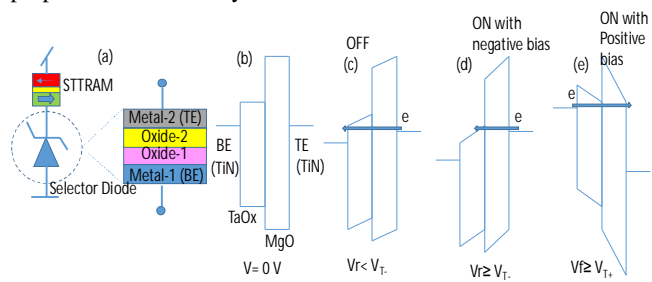

Fig. 2: (a) MIIM diode stack, (b) band-diagram at 0 bias, (c) band-diagram at negative bias (Vr) on TE ( Vr< VT-), (d) band-diagram at negative bias (Vr) on TE ( Vr> VT-), (e) band-diagram at positive bias (Vf) on TE ( Vf> VT+). BE is grounded.

In summary we make following contributions in this paper:
(a) We propose voltage biasing to enable multi-bit read operation and perform detailed analysis to support large crossbar size.
(b) Our analysis shows that a biasing of 0.7V can enable 512-bit read in 512x512 crossbar.
(c) We propose a voltage pulsing to enable multi-bit write operation and perform detailed analysis.
(d) We propose SD optimization to lower the energy-overhead of multi-bit write.
(e) The proposed techniques are also applicable to other crossbar arrays such as ReRAM and PCRAM.

The paper is organized as follows. Section II provides an overview of the MIIM SD [5] that is employed in this study. The proposed multi-bit read operation is described in Section III. The multi-bit write operation is described in Section IV. Conclusions are drawn in Section V.

## II. MIIM DIODE

Since STTRAM requires bidirectional current for switching, the SD has to be bidirectional i.e. it requires switching when applied bias is higher than $V_{T+}$ ($V_{T-}$) in positive (negative) bias along with low-reverse saturation current when applied bias is less than $|V_T|$. This has been studied and a MIIM SD has been proposed [5] (Fig. 2). The operation of the SD can be explained as follows: In forward bias (i.e. positive voltage on TE) the electron will experience the thickness of oxide-1 resulting in Direct Tunneling (DT) and when the bias voltage exceeds the positive threshold voltage ($V_{T+}$) Fowler-Nordheim (FN) current will dominate through the triangular barrier due to the lower barrier height of BE on HfO2 ($\Phi$BE) (Fig. 2(e)). This will turn-ON the device at $V_{T+}$.

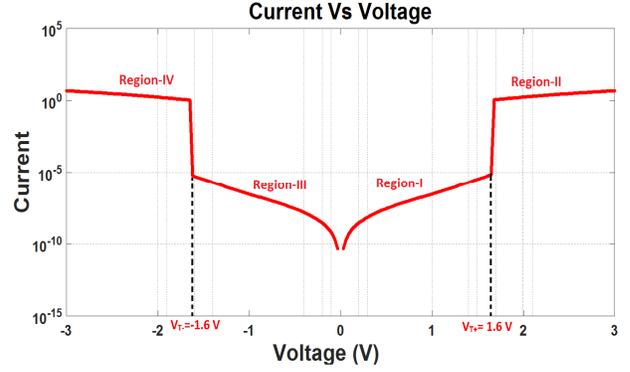

Fig. 3 I-V characteristics of MIIM SD.

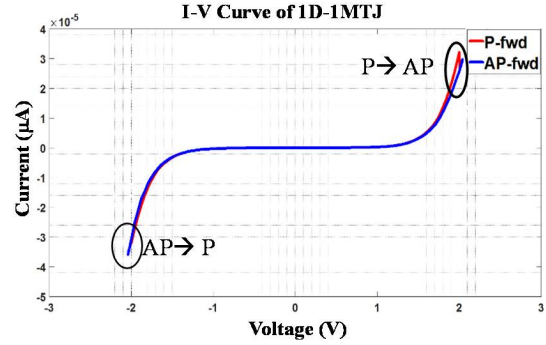

Fig. 4: 1D-1MTJ I-V curve for MTJ ($R_p$=6K, $R_{ap}$=12K)

The work functions of the metal-1 and metal-2, and the bandgap and electron affinities of oxide-1/oxide-2 can be selected to achieve the desired I-V characteristics. For example, Fig. 3 shows the I-V plot for TiN/TaOx/MgO/TiN MIIM diode. This is simulated using the sum of DT and FN currents equations shown below [5]:

$$J_{FN} = AE_{ox}^2 exp\left(-\frac{B}{E_{ox}}\right) \quad (1)$$

$$A = 1.54e-6\left(\frac{1}{m_{ox}/m_e}\right)^{1/2}\left(\frac{1}{\Phi_B}\right) \quad (2)$$

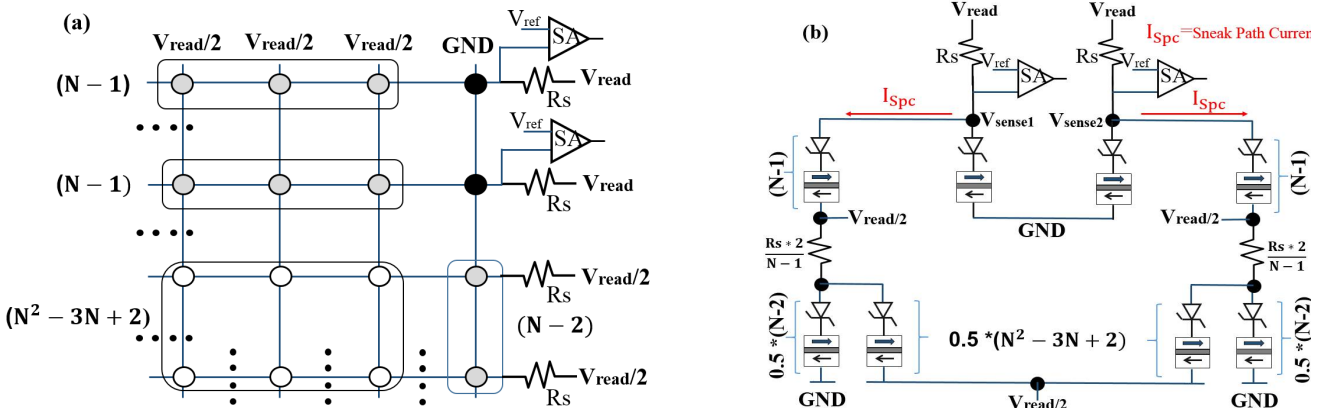

Fig. 5 (a) Proposed circuit for two bit read operation (shown with dark circles); and, (b) the sense voltages ($V_{sense1}$ & $V_{sense2}$) are a function of the two MJT that are read and the sneak path current through the half-selected bits and unselected bits. The sneak path equivalent circuit for two bit read is shown.

$$B = 6.8327e7\left(m_{ox}/m_e\right)^{1/2}(\Phi_B)^{3/2} \quad (3)$$

$$J_{DT} = AE_{ox}^2 \exp\left[\frac{-B\left(1-\left(1-\frac{V}{\Phi_B}\right)^{3/2}\right)}{E_{ox}}\right] \quad (4)$$

$$J_{total} = J_{FN} + J_{DT} \quad (5)$$

$$V_T = (\Phi_{TE-MgO} - \Phi_{MgO-HfO2})\left[\frac{\varepsilon_{MgO}t_{HfO2}}{\varepsilon_{HfO2}t_{MgO}} + 1\right] \quad (6)$$

Where, $J_{FN}$ and $J_{DT}$ are the Fowler-Nordheim and direct tunneling current densities respectively. A and B are diode constants, $E_{ox}$ is the electric field, $\Phi$ is the barrier height, $\varepsilon$ is the di-electric constant, $t_{ox}$ is the oxide thickness and $V_T$ is the threshold voltage.

The metal oxide to metal ($m_{ox}/m_0$) ratio of 0.26 is used in our simulations. Electron affinity of TaOx and MgO of 3.5 eV, and 2.3eV respectively [7], and, dielectric constant value of TaOx and MgO of 22, and 10 respectively, are used and the work-function of TiN of 4.0 eV is also incorporated. The I-V plot is divided into four regions. Region-I and Region-III is generated assuming tunneling through 1.5nm TaOx, Region-II and Region-IV is generated assuming tunneling through 0.1nm MgO. This unique characteristic is difficult to achieve using a single insulator with TE and BE of different work functions.

In this work we have assumed an ideal MIIM diode i.e. without bulk and interfacial defects and MTJ with Tunnel Magneto-Resistance (TMR) ratio = 2. Fig. 4 shows the I-V curve of 1D-1MTJ for $R_p$ (parallel state resistance) = 6KΩ and $R_{ap}$ (antiparallel state resistance) = 12KΩ. It is evident that the bit cell requires a write voltage of 2.1V.

## III. MULTI-BIT READ SCHEME

In this section we present the proposed memory cell and the methodology to read multi-bit data simultaneously.

### A. Design and Analysis

The proposed circuit for two bit read operation (Fig. 5 (a)) shows that when two bits are selected for read in NxN array, 2(N-1) + (N-2) bits are half-selected and ($N^2$-3N+2) bits are unselected. The design faces a few challenge during read operation. First, while reading two bit, the corresponding bits should be selected without disturbing other bits (half-selected and unselected bits). Even the two selected bits can be disturbed significantly if the read voltage (therefore, read current) is too high. First, each senseamp (SA) has to distinguish 2 states (0 and 1) for each memory cell. Therefore these two states per memory cell should have a maximum margin between them for robust sensing. Second, with the increase of crossbar array size, the sense margin shrinks due to sneak path current. Therefore, sneak path current need to be minimized for larger array for robust sense margin.

The voltage at $V_{sense1}$ and $V_{sense2}$ in Fig. 5 (b) is a function of TMR of the selected MJT and the sneak path current through the half-selected and unselected bits. Fig. 6 shows the sense margin with MTJ resistance ($R_p$). It is evident that on one hand sense margin increases with $R_p$. But on the other hand, higher $R_p$ also requires higher write current i.e. higher write voltage. We have chosen $R_p$=6KΩ for optimum sense margin and $V_{write}$.

The sense resistance ($R_s$) also plays a vital role in determining the sense margin. Fig. 7 shows the sense margin with respect to $R_s$ for 50x50 crossbar array. The sense margin increases till

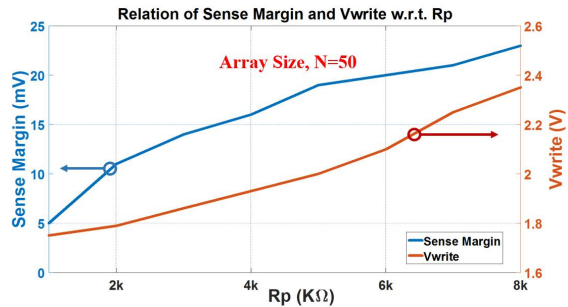

Fig. 6 Relation of 2-bit sense margin and Vwrite w.r.t. Rp for N=50

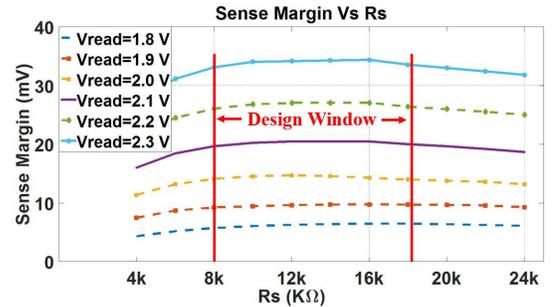

Fig. 7 Sense margin of 2-bits vs Rs for array size, N=50

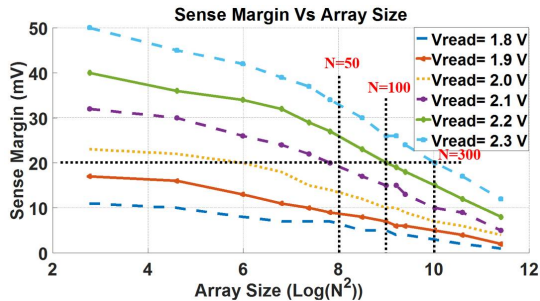

Fig. 8 Sense margin of 2-bits vs array size with Rs= 16KΩ

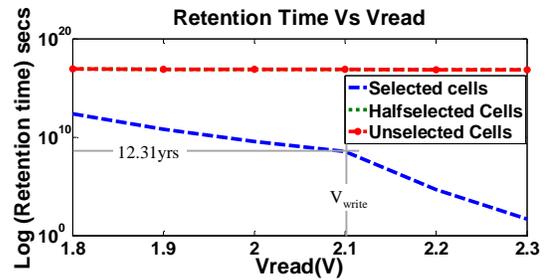

Fig. 9 Retention time of 2-bits vs $V_{read}$ for Rs=16KΩ and N=50.

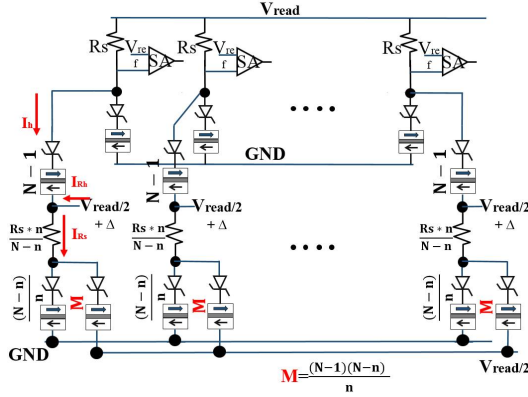

Fig. 10 Equivalent circuit for n-bit read operation for a N*N array

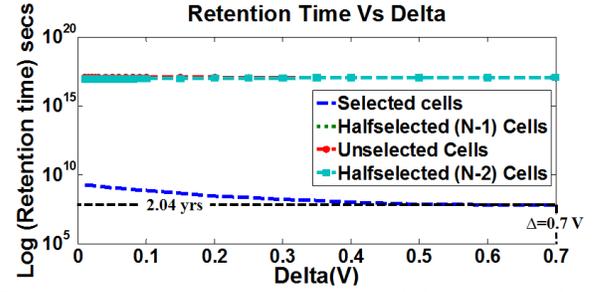

Fig. 11 Retention time vs delta with $V_{read}$=2.1V and N=50

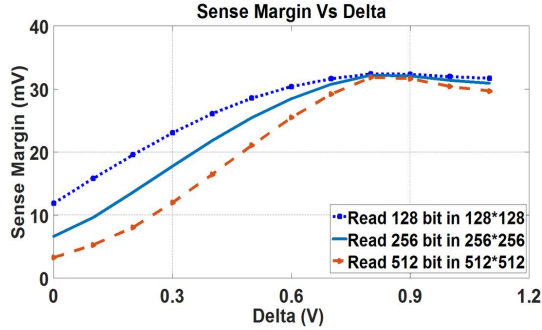

Fig. 12 Sense margin vs delta with $V_{read}$=2.1V

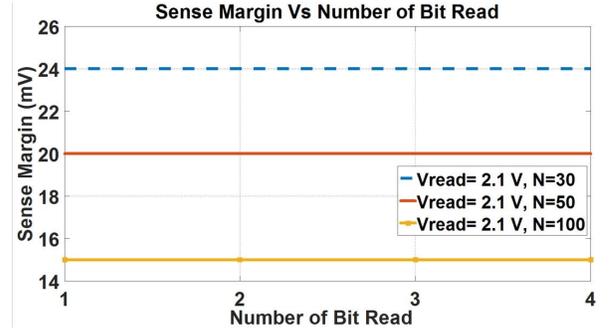

Fig. 13 Sense margin for multiple bit read operation

18KΩ but decreases afterwards. Therefore, the design window for $R_s$ is 8KΩ to 18KΩ. Higher $R_s$ ensures lower read current and lower sneak path current. Therefore in this work we have chosen $R_s$ =16KΩ.

*B. Read and Retention*

As mentioned earlier, sneak path current is a design limitation as the array size increases. To incorporate all the possible path for sneak current and simulate the read operation for two bits in NxN crossbar array, an equivalent circuit has been developed which is shown is Fig. 5(b). For calculating the worst case sense margin, all the MTJs in the sneak path has been considered of having the low resistance.

The circuit is simulated for different $V_{read}$ values and crossbar array sizes. It is evident from simulation result (Fig. 8) that 20mV sense margin can be achieved for array size N=50 @$V_{read}$=2.1V, N=100 @$V_{read}$=2.2V and N=300 @$V_{read}$=2.3V. The retention time for selected cells, half selected cells and unselected cells have been presented in Fig. 9. Although higher $V_{read}$ gives better sense margin for larger array, the retention time goes down with $V_{read}$. The thermal energy of the MTJ used in the analysis is $60k_bT$, where $k_b$ is the Boltzmann constant and T is the absolute temperature. The volume of the MTJ is assumed as $2e^{-18}$ cm$^3$. In this work, we have chosen $V_{read}$ = 2.1V. Although $V_{read}$ is same as $V_{write}$, the selected cells get voltage below 1.9 V due to Rs and drop across SD and therefore they are not written. At $V_{read}$= 2.1V, the selected cells have around 12.31 years of retention time.

To increase the crossbar array size and perform read operation smoothly the sense margin needs to be improved. We propose to use $V_{read}/2+\Delta$ instead of $V_{read}/2$ to bias all unselected bit lines and word lines, and, improve the sense margin, where Δ is a small positive voltage. Fig. 10 represents an equivalent circuit for n-bit read operation in NxN array. The current $I_h$ lowers the sense margin for larger array. If Δ is increased the current $I_h$ decreases and the sense margin improves. However, increasing Δ will in turn increase $I_{Rs}$. Therefore, some half-selected cells will have improved retention time whereas some other half-selected cells will have lower retention time. Retention time of selected and half selected and unselected cells are presented in Fig. 11. It is evident from the result that retention time for the selected cells have decreased for higher Δ. However, the retention time for (N-1) half-selected has improved and retention for (N-2) half-selected has decreased. Fig. 12 shows that as Δ increases the sense margin improves and reaches a saturated value of 32 mV which is the sense margin for 2 bit read in 4x4 array. The retention time at selected Δ=0.7 is 2.04 years.

*C. Multi-bit Read*

Fig. 13 shows the sense margin for multi-bit read operation. For a NxN array if any n-bits are read (where n=2, 3, .., N), the sense margin remains same. As shown in Fig. 10, the sense margin is mainly affected by the (N-1) half-selected cells that are in the same row of the selected cell in NxN crossbar array. The other half-selected and unselected are equally distributed among the selected cells (Fig. 10). With higher Δ, $I_h$ decreases and $I_{Rs}$ increases and since $I_{Rh} = I_{Rs} + I_h$, the extra current is adjusted by $I_{Rh}$. If N bits are read from NxN array and $V_{read}/2+\Delta$ are applied to all bit lines and word lines, the sense margin can be increased up to 32mV (for $\Delta$=0.7V) and the retention of (N-1) half-selected will also be maximized (Fig. 10). From our analysis we conclude that it is possible to sustain as much as 512 bit read with $\Delta$=0.7V in 512x512 crossbar (Fig. 12). The retention time is found to be 2.04 years.

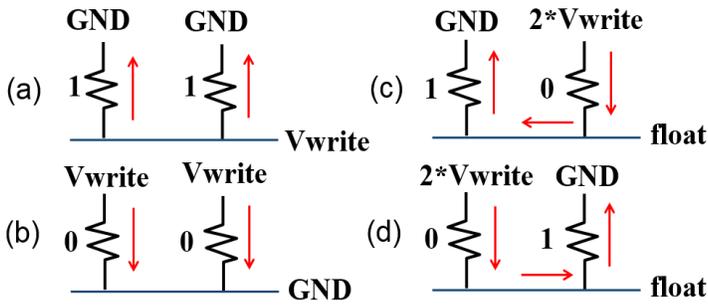

Fig. 14 Two bit writing simultaneously for four combinations

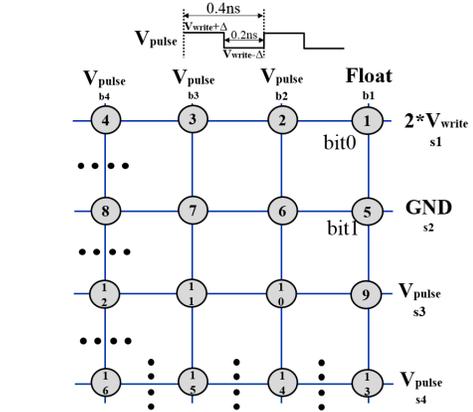

Fig. 15 Writing '10' with 2*$V_{write}$ (4.2V)

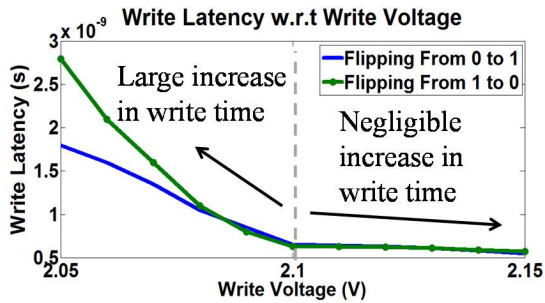

Fig. 16 Write latency w.r.t. write voltage

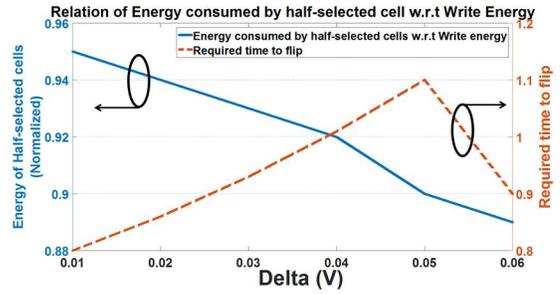

Fig. 17 Energy consumed by half-selected cells and time to flip

## IV. MULTI-BIT WRITE SCEME

In this section, we discuss the methodology to write 2-bits of memory simultaneously. Writing 2-bit together can be categorized into two group: (a) writing identical bits i.e. '11' or '00'; and, (b) writing complementary bits i.e. '10' or '01'. Both of these cases are explained in this section.

### A. Design and Analysis

Fig. 14 represents writing two bits simultaneously. For writing '11' (Fig. 14 (a)), we apply $V_{write}$ to the bit line and ground the word line of the selected bits. For writing '00' (Fig. 14 (b)), we apply $V_{write}$ to the word lines and ground the corresponding bit lines. For ensuring the bits are not disturbed $V_{write}/2$ is applied to all other word and bit lines. Therefore writing '00' and '11' is straightforward. However, the challenge arises for writing complementary bits (Fig. 14 (c) and Fig. 14 (d)) when 2 SDs and 2 MTJs are in series. The write voltage should be 2$V_{write}$ to write both bits and the unselected word lines and bit lines should be $V_{write}$. The selected bit line is floated. However, this will end up writing the half-selected bits.

From Fig. 15 it is evident that applying $V_{write} + \Delta$ and $V_{write} - \Delta$ on unselected lines will prevent one set of half-selected bits (Cell 2, 3, 4) from getting written whereas the other set (Cell 6,7,8) will be written. To solve this issue we apply a pulse voltage that varies between $V_{write} + \Delta$ and $V_{write} - \Delta$ (Fig. 15) to all unselected bit lines and word lines. We note that the MTJ needs long time to flip from 0 to 1 or 1 to 0 for a specific write voltage. Furthermore the write delay increases non-linearly at lower write voltages whereas it increases linearly at higher write voltage (Fig. 16). Therefore by choosing $V_{write}$ and $\Delta$ carefully we prolong the write latency of half-selected cells (Cells 2, 3, 4 and 6, 7, 8) so that they stay undisturbed by the time the selected cells (Cells 1 and 5) are written. Fig. 16 shows the simulation results for write latency with respect to $V_{write}$. For $V_{write} < 2.1$ V, the write latency reduced rapidly however, for $V_{write} > 2.1$ V, write latency stays almost same. In this work, $\Delta$ is chosen as 0.05V with $V_{write} = 2.1$ V. Although, the half-selected cells suffer significantly in terms of retention (~15ns) the 2-bits are written successfully.

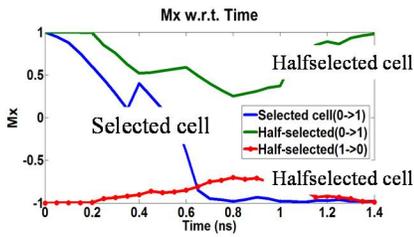

Fig. 18 Magnetization (Mx) of selected and half-selected cells with respect to time.

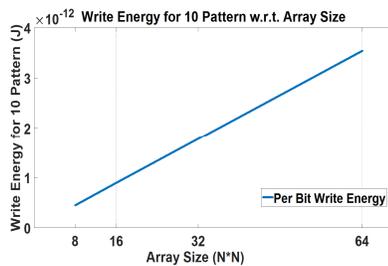

Fig. 19 Write energy for '10' pattern with respect to size of crossbar

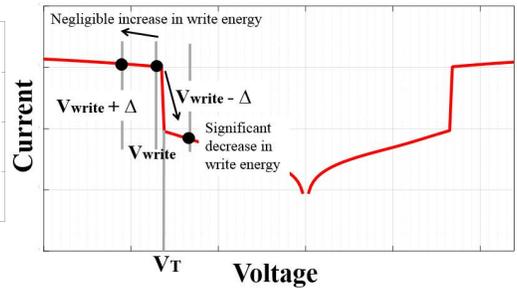

Fig. 20: $V_T$ modulation of SD to reduce leakage current of half-selected bits.

*B. Simulation Results*

In Fig.18, magnetization (Mx) with respect to time is shown with $V_{pulse}$ applied to the half-selected cells and $2V_{write}$ to selected cells. It is evident that ~0.75 ns is required to flip both cell together (cell 1 and 5 in Fig. 15) with $2V_{write} = 4.2$V. However, the half-selected cells go back to their correct states when the biasing voltage becomes zero after 0.8ns. If the system clock speed is considered to be 1.25GHz, 1 clock cycle is needed for this operation. However, since the half-selected cells get a pulse voltage their write time is higher. In this work, the pulse width of 0.4 ns with 50% duty cycle is applied to unselected word lines and source lines. The pulse toggles between 2.15 V and 2.05 V. Simulation result shows that with the pulse voltage source the flipping time from 0 to 1 becomes 1.1 ns and flipping time from 1 to 0 becomes 1.59 ns. If the worst case is considered, 1.5 clock cycles will flip the half-selected cells. Therefore, same crossbar cannot be accessed back to back. A time gap between two consecutive accesses is required in the crossbar so that the half-selected cells relax and become stable again. This can be taken care at the bank level by prohibiting back-to-back writes to a bank.

A shortcoming of the proposed write technique is energy penalty while writing '10' and '01'. If the size of the crossbar array is NxN, the current through each of the 2(N-1) half-selected cells will be almost equal to the write current. Therefore, these cells will consume significant energy (Table 1 for 64x64 crossbar array). Note that although the energy overhead of 10/01 pattern in high a cache line will contain only a fraction of such patterns to partially mitigate the overhead. We propose following techniques to lower the energy overhead further:

(a) Adjustment of the value of $\Delta$ and adjustment of $V_T$ of the MIIM diode. The I-V curve of diode in Fig. 3 shows that $V_{write}-\Delta$ reduces the current non-linearly compared to the current increment by $V_{write}+\Delta$. Fig. 17 shows the average energy of the half-selected cells with respect to $\Delta$. The consumed energy is reduced to 5% by varying delta.

Table 1: Write energy vs bit pattern for 64x64 crossbar

| Data | Write energy in pJ (per bit) |
|---|---|
| 00/11/10/11 | 0.089/0.091/3.53/3.53 |

(b) Reducing the crossbar size. The per bit writing energy decreases as the number of half-selected cells (in s1 and s2 row of Fig. 15) decreases (Fig. 19). Therefore smaller crossbar can lower the energy overhead.

(c) If the $V_T$ of the MIIM diode is modulated in a way that the 1D-1MTJ writes at $V_{write}$ but at $V_{write} - \Delta$ the diode turns off (very low current through the cell) and at $V_{write} + \Delta$, current through the cell is not very high compared to $I_{write}$ (Fig. 20), one row of half-selected cells remains turn off. Therefore, 50% write energy saving for writing 10 or 01 is possible.

(d) By encoding the cache line data to increase the population of 00/11 compared to 01/10.

V. CONCLUSION

In this work two bit read and write schemes have been presented. We improve the sense margin during read to sustain larger crossbar arrays. We propose voltage biasing technique for multi-bit read operation. We also propose two-bit write by employing pulsed voltage on unselected word line and bit lines.